\documentclass[12pt,preprint]{aastex}

\slugcomment
{ApJ, in press, April 1, 2007 issue}

\begin{document}

\title{The 15 -- 20 Micron {\it Spitzer} Spectra of
Interstellar Emission Features in NGC 7023}
 
\author{
K. Sellgren\altaffilmark{1},   
K. I. Uchida\altaffilmark{2},   
and M. W. Werner\altaffilmark{3}   
}

\affil{}
 
\email{
sellgren@astronomy.ohio-state.edu
}

\altaffiltext{1}
{Department of Astronomy, Ohio State University, 140 W. 18th Av.,
Columbus, OH 43235}
\altaffiltext{2}
{Jet Propulsion Laboratory, MS 301-465, 4800 Oak Grove Drive, Pasadena, 
CA 91109}
\altaffiltext{3}
{Jet Propulsion Laboratory, MS 264-767, 4800 Oak Grove Drive, Pasadena, 
CA 91109}

\clearpage

\begin{abstract}
We present 15 -- 20 $\mu$m long-slit spectra, from the
Infrared Spectrograph (IRS)
on {\it Spitzer}, of NGC 7023.  
We observe 
recently-discovered interstellar emission features, at 
15.9, 16.4, 17.0, 17.4, 17.8, and 18.9 $\mu$m,
throughout the reflection nebula.
The 16.4 $\mu$m emission feature
peaks near the photodissociation front
northwest of the star,
as do the aromatic emission features (AEFs) at 
3.3, 6.2 and 11.3 $\mu$m.
The 16.4 $\mu$m emission feature is thus likely related to the AEFs
and radiates by non-equilibrium emission.
The new 18.9 $\mu$m emission feature,
by contrast,
decreases monotonically with 
stellar distance.
We consider candidate species for
the 18.9 $\mu$m feature,
including polycyclic aromatic hydrocarbons,
fullerenes,
and diamonds.
We describe future laboratory and
observational research needed to
identify the 18.9 $\mu$m feature carrier.
\end{abstract}


\clearpage
\section{Introduction}
The spectrum of aromatic emission features 
(AEFs) between 3.3 
and 12.7 $\mu$m has been previously well characterized
by ground-based, airborne, and space-based spectroscopy.
The 15 -- 20 $\mu$m region of AEF sources, however, has
been far less examined.
\citet{Beintema96} obtained
2.4 -- 45 $\mu$m 
{\it Infrared Space Observatory (ISO)} 
spectra of three bright AEF sources,
using the Short Wavelength Spectrograph (SWS).
They found new emission features in the 
15 -- 20 $\mu$m region at
17.0 (broad) and 17.9 $\mu$m.
\citet{Moutou00}, \citet{VanKerckhoven00},
and \citet{Sturm00} discovered a
new emission feature at 16.4 $\mu$m in a variety of
sources, with {\it ISO}-SWS spectra.
The 16.4 $\mu$m feature was identified with C--C--C skeletal
vibrational modes in polycyclic
aromatic hydrocarbon (PAH) molecules with pentagonal
rings \citep{Moutou00} or the C--C--C in-plane bending of 
PAHs with pendant hexagonal rings \citep{VanKerckhoven00}.
\citet{Sturm00} discovered
a new 15.8 $\mu$m emission feature in {\it ISO}-SWS spectra
of five star-forming galaxies.
\citet{VanKerckhoven00} detected a broad
plateau of emission at 15 -- 20 $\mu$m in several sources, 
corresponding to the broad 17.0 $\mu$m feature discovered by 
\citet{Beintema96}.

The Infrared
Spectrograph (IRS; \citealt{IRS}) on {\it Spitzer} \citep{SIRTF}
has the sensitivity to detect significant 
spectral substructure in
the 15 -- 20 $\mu$m region not evident in previous observations.
\citet{WUS04} (hereafter
\defcitealias{WUS04}{Paper~1}
\citetalias{WUS04}) 
have discovered new spectral
features at 17.4 and 19.0 $\mu$m,
in addition to previously detected narrow 
features at 15.8, 16.4 and 17.8 $\mu$m
and a broad feature at 17.0 $\mu$m,
in {\it Spitzer} 
IRS spectra of \objectname[]{NGC~7023}.
We describe in \S 3.3 how we use higher
resolution spectroscopy to improve the wavelengths
for these features; as a result, the `15.8' and `19.0' $\mu$m
features become the `15.9' and `18.9' $\mu$m features. 
These spectral features at 15.9, 16.4, broad 17.0,
17.4, 17.8, and 18.9 $\mu$m have now all been observed
in {\it Spitzer} IRS spectra
of other star formation regions and galaxies
(\citealt{Armus04}; 
\citealt{Brandl04}; 
\citealt{Morris04}; 
\citealt{Smith04}; 
\citealt{Higdon06}; 
\citealt{Engelbracht06}; 
\citealt{Tappe06};
\citealt{Panuzzo06};
\citealt{Smith06}),
including the new 18.9 $\mu$m feature which
has been detected in several galaxies
(\citealt{Panuzzo06}; \citealt{Smith06}).

We extend here the work of
\citetalias{WUS04}, 
by presenting additional 
{\it Spitzer} IRS spectroscopy of NGC 7023,
a well-studied photodissociation region (PDR) and reflection nebula.
NGC 7023 is illuminated by the Herbig Be star 
\objectname[]{HD~200775},
at a distance of 430 pc \citep{vdA97}.
This extension of our previous work allows us to better
characterize the spatial distributions 
and emission mechanisms of the features
in the 15 -- 20 $\mu$m region.

\section{Observations}

We describe our observational technique in \citetalias{WUS04}.
We improve the data reduction and examine here, in further detail,
the Short-High (SH) IRS spectrum 
(9.9 -- 19.6 $\mu$m; $R$ = $\lambda$/$\Delta \lambda$ = 600) 
of NGC 7023 at one pointing position previously presented in
\citetalias{WUS04}, obtained on 2003 September 19.
The SH integration time was 6 s, repeated twice.
We extract the SH spectrum from 
the entire aperture (4\farcs7 $\times$ 11\farcs3).
We reduce 
three new slit pointing positions (24 additional spectra in NGC 7023) of
second-order Long-Low (LL2) long-slit IRS spectra 
(14.0 -- 21.3 $\mu$m; $R$ = 80 -- 128), obtained on 2003 October 1.
The LL2 integration time was 30s, repeated 3 times.
We refer to these new slit pointing positions as slits A, C, and D.
In this notation, the LL2 slit in \citetalias{WUS04}
is slit pointing position B.

\begin{figure*}
\figurenum{1}
\includegraphics[width=6.45in]
{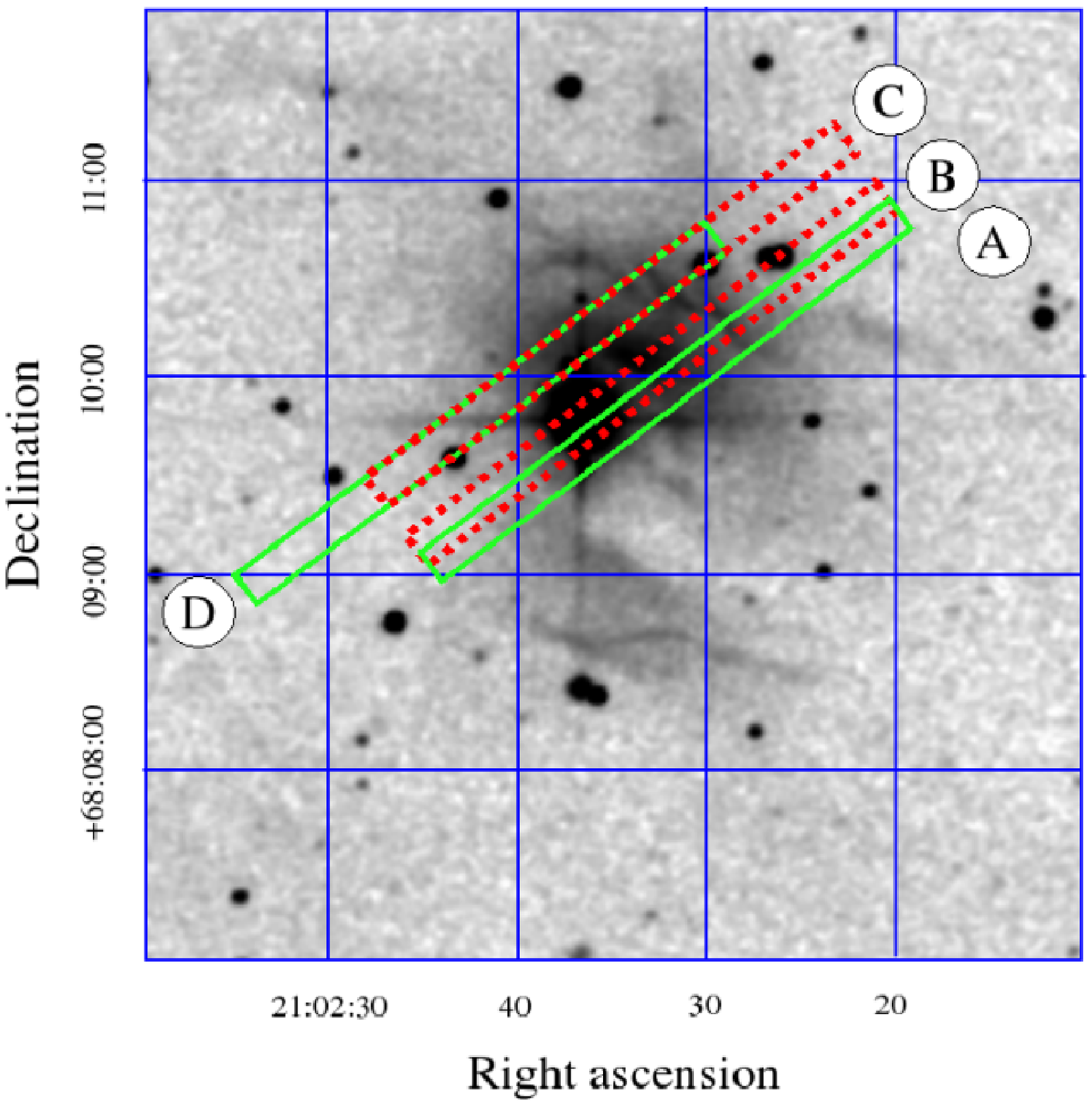}
\caption{
We illustrate how the IRS
slit (10\farcs5 $\times$ 168\arcsec)
at each pointing position (slit A, B, C, or D)
overlays different parts 
of NGC 7023 on a 2MASS 
\citep{2MASS}
K$_{s}$ image.
We extract LL2
spectra at eight spatial positions from each long-slit,
with position 1 furthest southeast and position 8
furthest northwest.
Offsets from HD 200775 range from 72\arcsec\ east 30\arcsec\ south
(slit D, position 1) to
50\arcsec\ west 61\arcsec\ north (slit C, position 8).
Slit B, position 3 is HD 200775.
}
\label{fig_n7023_slits}
\end{figure*}

Figure \ref{fig_n7023_slits} illustrates the location
of slit pointing positions A, B, C, and D.
We extract eight spectra from each slit, at
evenly-spaced spatial positions we refer to as positions 1
through 8, labeled from southeast to northwest
(see Fig. \ref{fig_n7023_slits}).
All of the slits A, B, C, and D were observed
at a slit position angle 306.4$\degr$ east of north.
Slit B crosses HD 200775 at position 3 (spectrum L3 in 
\citetalias{WUS04}).
Slits A and C are shifted relative to B by 6\farcs4 southwest
and 19\farcs2 northeast, respectively, 
in a direction perpendicular to the slit.
Slit D is shifted 
southeast by 45$''$, relative to C, in a direction parallel
to the slit.
We extract spectra from fixed pixel ranges 
(extraction box of 2 $\times$ 3 pixels, 
or 10\farcs2 $\times$ 15\farcs3) along the 
2 pixel wide slit
of LL2, producing eight spectra for each slit pointing position.
We re-reduce Slit B, both for consistency with the other
spectra presented in this paper and because of improvements in
the calibration;
this changes the intensities
of Slit B spectra
by $\sim$7\%.
We present here an analysis of a total of 32 spectra.
All of our data was obtained prior to nominal {\it Spitzer}
operations (pre-nominal status), so our intensity, wavelength,
and pointing uncertainties are larger than for nominal
(current) {\it Spitzer} operations.
The SH and LL2 absolute positions are thus uncertain by $\pm$5\arcsec.

We reduce the data and extract spectra
using Cornell IRS 
Spectroscopy Modeling Analysis and Reduction Tool
\citep{Higdon04}.  
All data are hand-calibrated by applying the spectrum and model 
template of point source calibrator 
\objectname[]{HR~8585}
($\alpha$ Lac)
to the intermediate 
un-flat-fielded product of the SSC pipeline. 
We subtract blank sky spectra only 
from the SH spectrum, observed as {\it Spitzer} cooled,
to remove thermal emission from the then warm ($\sim$45 K) baffles 
of the telescope. 
We apply the current extended source correction
to the LL2 spectra.
The LL2 spectra have an absolute uncertainty in their
intensities
of $\sim$15\%, due to their
being obtained very early in the {\it Spitzer} mission,
before the detector bias was changed.

Figure \ref{fig_n7023_slits} 
shows that three of the four LL2 slits we observe
intersect the filaments seen to the northwest of HD 200775.
The filaments, seen in Figure \ref{fig_n7023_slits},
trace a PDR front \citep{Lemaire96}.
All four LL2 slits
also cover low-intensity regions to the southeast of HD 200775.
The LL2 slit illustrated in \citetalias{WUS04}
is shown in Figure \ref{fig_n7023_slits} as Slit B, and it
intersects both the northwest filaments and HD 200775.
The SH spectrum \citepalias{WUS04} is obtained 0$''$W 29$''$N
of HD 200775, between the star and the PDR front.

\section{Results}

\subsection{Observed Features at High Resolution}

\begin{figure*}
\figurenum{2}
\includegraphics[width=6.50in]
{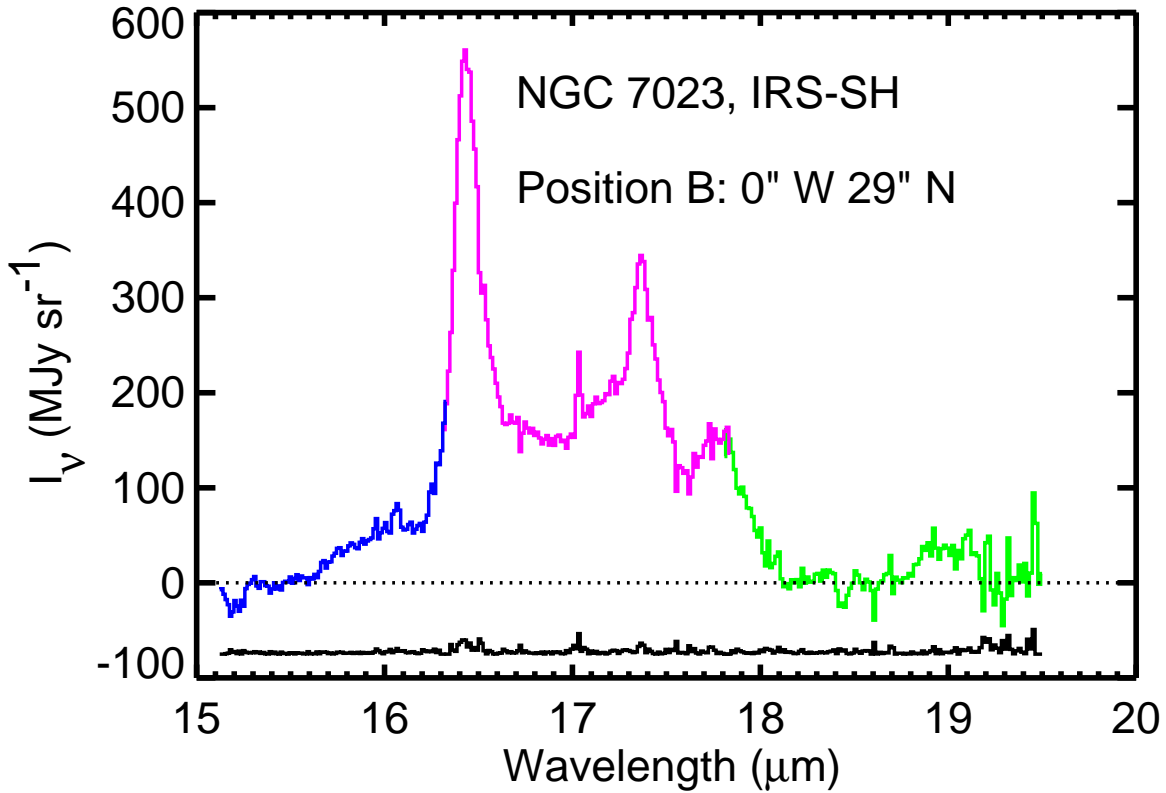}
\caption{
Continuum-subtracted SH spectrum
(15.1 -- 19.5 $\mu$m; $R$ = 600)
of Position B
(0\arcsec W 29\arcsec N; \citetalias{WUS04}) in NGC 7023.
We measure the intensity $I_\nu$ (MJy sr$^{-1}$) 
from the entire entrance
slit (4\farcs7 $\times$ 11\farcs3).
We illustrate three overlapping SH orders, in contrasting colors.
The wavelengths of individual orders
are shifted by $\leq$ 0.05 $\mu$m to
match adjacent orders.
The zero point for the wavelength scale is set
by the 0--0 S(1) line at 17.03 $\mu$m.
We plot below the spectrum, 
as a measure of our uncertainty,
the root-mean-square (rms) difference
between the spectrum and a
7-point moving boxcar average of the spectrum.
}
\label{fig_sh}
\end{figure*}

Figure \ref{fig_sh}
illustrates our SH spectrum
(15.1 -- 19.5 $\mu$m; $R$ = 600) in NGC 7023.
This is a part of the SH spectrum shown in
\citetalias{WUS04}, although we improve the
reduction, better matching the spectral segments
by shifting wavelength zero points in each order 
by $\leq$ 0.05 $\mu$m.
For these pre-nominal {\it Spitzer}
data, the SH wavelengths are more uncertain 
than relative intensity levels.
We establish an absolute wavelength scale 
with the 17.03 $\mu$m 0--0 S(1) H$_2$ line.
We also improve the contrast of weak features by
subtracting a continuum.

We plot uncertainties in
Figure \ref{fig_sh}
calculated from the root-mean-square 
difference between the data and a seven-point moving boxcar
average of the data.
The statistical uncertainties, as measured from the
difference between two individual integrations, are
too small to be visible as error bars.

Figure \ref{fig_sh} clearly shows resolved
emission features at 16.4, 17.4, and 17.8 $\mu$m. 
We detect weaker resolved emission features at 15.9 and 18.9 $\mu$m.
The broadness of these emission features 
(0.1 -- 0.6 $\mu$m, or 5 -- 20 cm$^{-1}$)
implies a solid-state material, a molecular band, or a 
molecular vibrational mode 
broadened for the same reasons that the AEFs are broad.
The only unresolved emission line in Figure \ref{fig_sh} is
the 0 -- 0 S(1) H$_2$ line at 17.03 $\mu$m.

\subsection{Long-Slit Spectra}

\begin{figure*}
\figurenum{3}
\includegraphics[height=7.10in]
{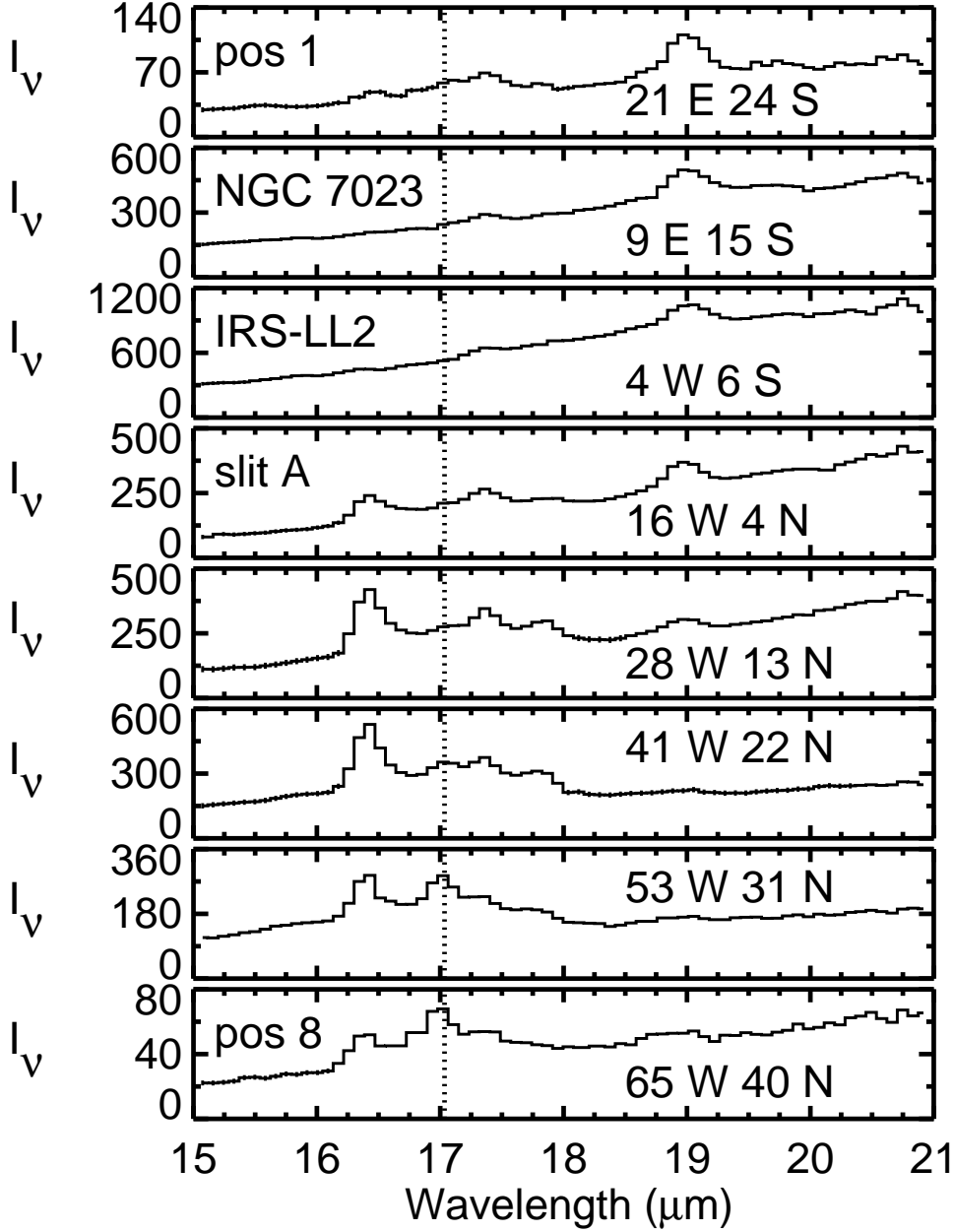}
\caption{
The 15 -- 21 $\mu$m LL2 spectra 
($R$ = 80 -- 128) of slit A.
We measure the intensity $I_\nu$ (MJy sr$^{-1}$) with an
extraction box of 10\farcs2 $\times$ 15\farcs3.  
We label each spectrum with its
offset (\arcsec) from HD 200775.
We mark the wavelength of the 0--0 S(1) H$_2$ line {\it (dotted 
line)}.
We plot $\pm$1-rms uncertainties, from the difference
between the spectrum and a
5-point moving boxcar average of the spectrum.
}
\label{fig_b15}
\end{figure*}

\begin{figure*}
\figurenum{4}
\includegraphics[height=6.65in]
{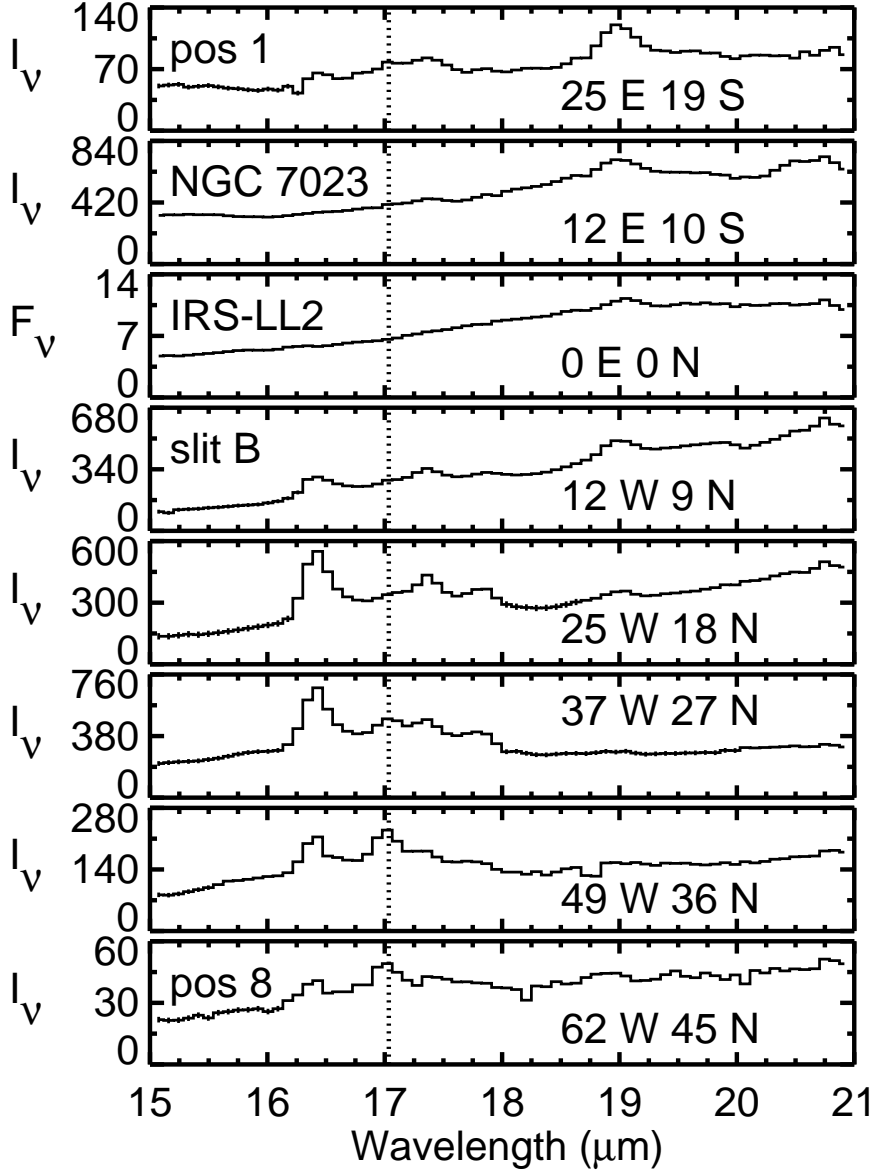}
\caption{
The 15 -- 21 $\mu$m LL2 spectra 
($R$ = 80 -- 128) of slit B.
We measure the intensity $I_\nu$ (MJy sr$^{-1}$) with an
extraction box of 10\farcs2 $\times$ 15\farcs3.
These spectra from \citetalias{WUS04} are
re-calibrated (change of $\sim$7\%) for consistency 
and shown here for comparison.
We label each spectrum with its
offset (\arcsec) from HD 200775.
The spectrum of HD 200775 (offset 0\arcsec east 0\arcsec north)
is in units of $F_\nu$ (Jy).
We mark the wavelength of the 0--0 S(1) H$_2$ line {\it (dotted 
line)}.
We plot $\pm$1-rms uncertainties, from the difference
between the spectrum and a
5-point moving boxcar average of the spectrum.
}
\label{fig_b17}
\end{figure*}

\begin{figure*}
\figurenum{5}
\includegraphics[height=6.65in]
{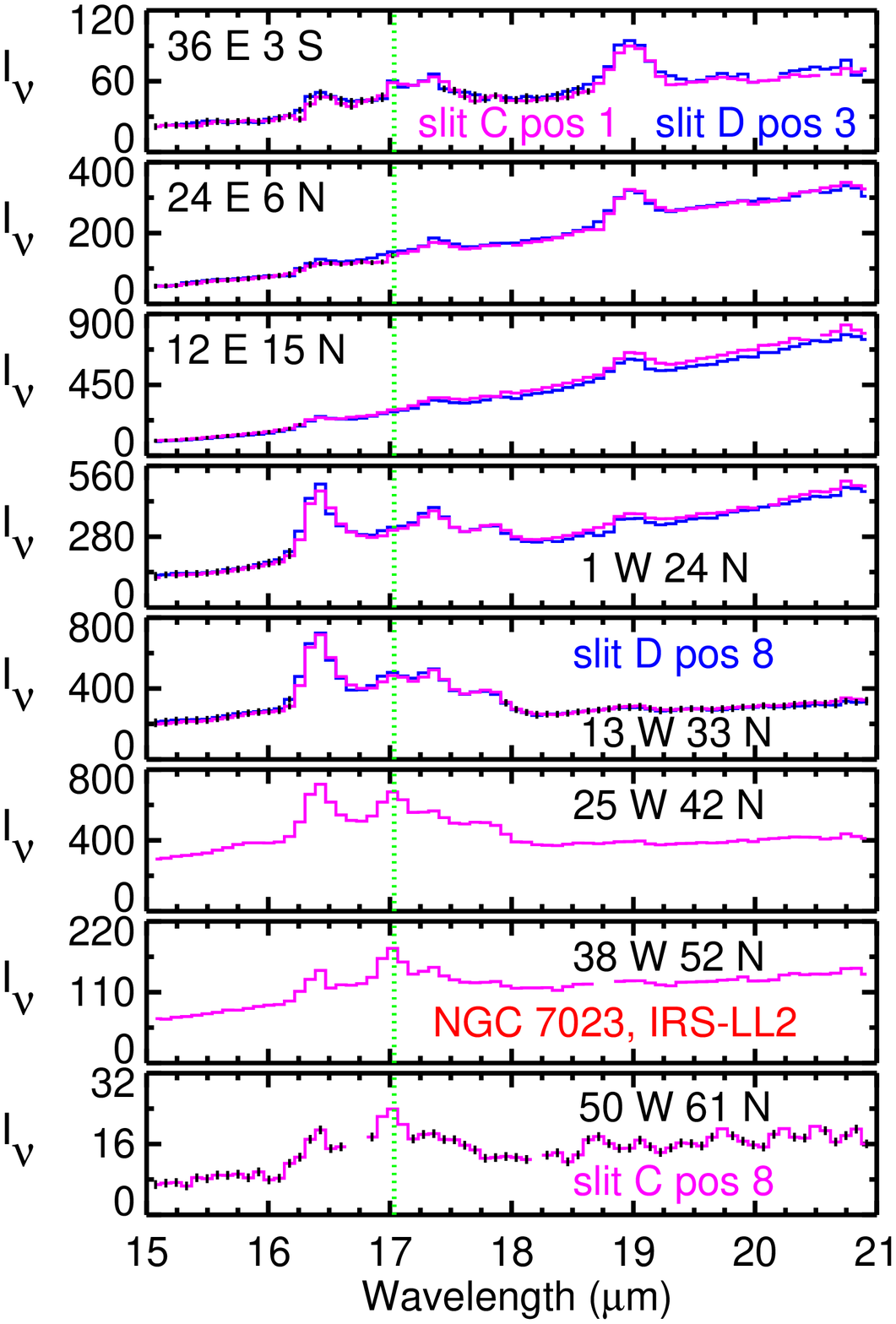}
\caption{
The 15 -- 21 $\mu$m LL2 spectra ($R$ = 80 -- 128) of slit
C {\it (magenta)}.
We measure the intensity $I_\nu$ (MJy sr$^{-1}$) with an
extraction box of 10\farcs2 $\times$ 15\farcs3.
Where there is spatial coincidence to within $\sim$1 pixel
($\sim$5\arcsec) between
positions in slits C and D,  
we have overplotted the slit D spectrum 
{\it (blue)}
on the slit C spectrum.
We label each spectrum with its
offset (\arcsec) from HD 200775.
We mark the wavelength of the 0--0 S(1) H$_2$ line 
{\it (dotted green line)}.
We plot $\pm$1-rms uncertainties, from the difference
between the spectrum and a
5-point moving boxcar average of the spectrum.
}
\label{fig_b23}
\end{figure*}

\begin{figure*}
\figurenum{6}
\includegraphics[height=6.65in]
{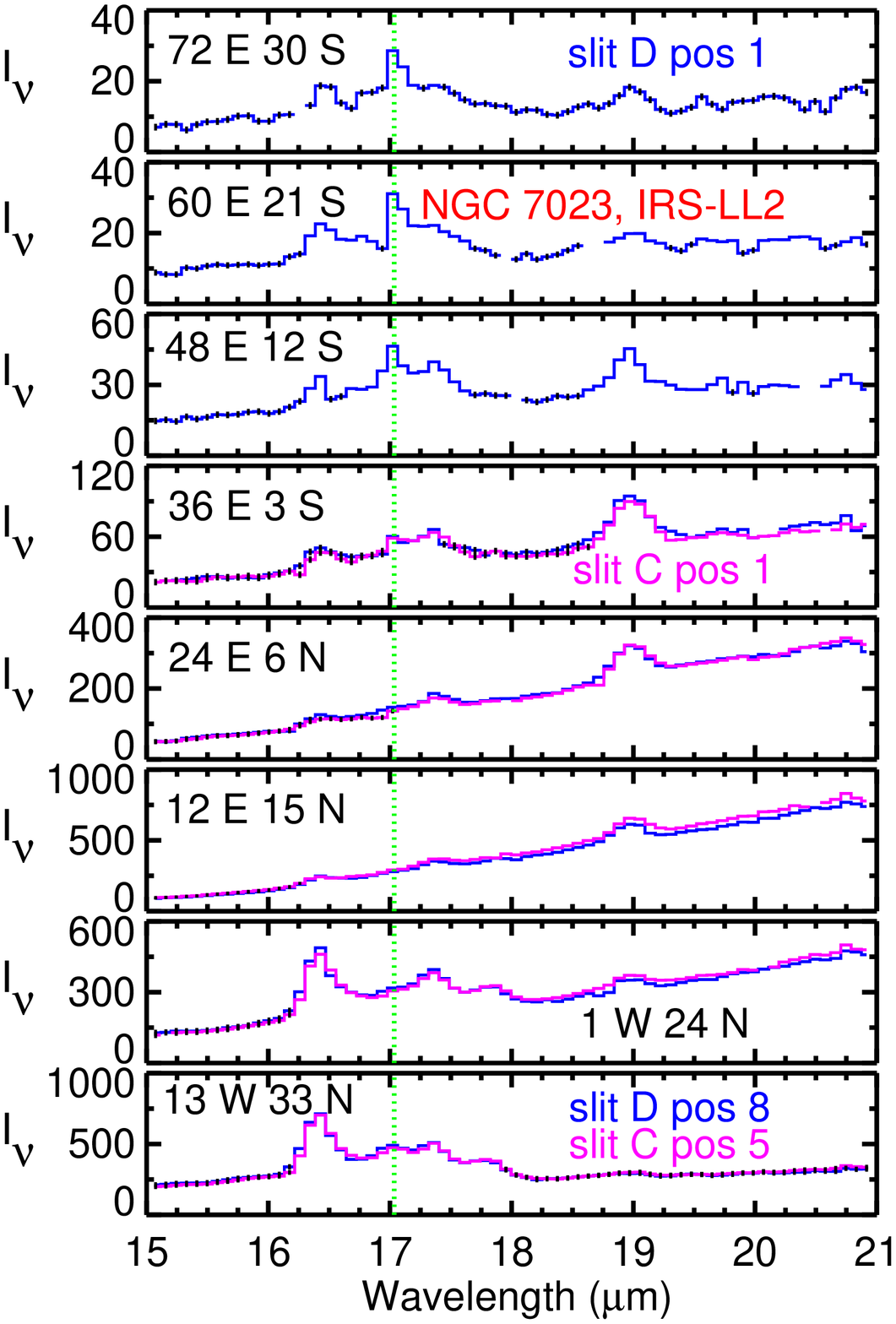}
\caption{
The 15 -- 21 $\mu$m LL2 spectra ($R$ = 80 -- 128) of slit
D {\it (blue)},
We measure the intensity $I_\nu$ (MJy sr$^{-1}$) with an
extraction box of 10\farcs2 $\times$ 15\farcs3.
Where there is spatial coincidence to within $\sim$1 pixel
($\sim$5\arcsec) between
positions in slits C and D,  
we have overplotted the slit C spectrum 
{\it (magenta)}
on the slit D spectrum.
We label each spectrum with its
offset (\arcsec) from HD 200775.
We mark the wavelength of the 
0--0 S(1) H$_2$ line {\it (dotted green line)}.
We plot $\pm$1-rms uncertainties, from the difference
between the spectrum and a
5-point moving boxcar average of the spectrum.
}
\label{fig_a6}
\end{figure*}

We present our LL2 spectra (14 -- 21 $\mu$m; $R$ = 80 -- 128)
in NGC 7023 in
Figures \ref{fig_b15}, \ref{fig_b17}, \ref{fig_b23}, and \ref{fig_a6}.
We plot uncertainties
derived in the same way as for Figure \ref{fig_sh},
except that a five-point moving boxcar was used.

Figure \ref{fig_b15} illustrates our new results for slit A.
These spectra demonstrate marked spectral variability from 
the PDR traced by the northwest filaments, dominated by the 16.4 $\mu$m
emission feature, the broad 17.0 $\mu$m feature, and 
the 0 -- 0 S(1) H$_2$ line at 17.0 $\mu$m;
to the regions between the PDR and the star,
where the 16.4, 17.4, and broad 17.0 $\mu$m
features are strongest;
to the southeast, where the 18.9 $\mu$m
feature is most prominent and
the 16.4 $\mu$m feature is extremely faint.

Figure \ref{fig_b17} shows our results for
slit B, from \citetalias{WUS04}, 
which have been re-calibrated ($\sim$7\% change)
and are illustrated here for comparison.
The spectral features vary with spatial position in slit B 
much as they do in slit A.
The 17.4 $\mu$m feature is most easily seen in spectra
where both the 16.4 and 18.9 $\mu$m
features are present.

Figures \ref{fig_b23} and \ref{fig_a6} show our new data for slits
C and D, respectively.
Figures \ref{fig_b23} and \ref{fig_a6} reveal the 
16.4 and broad 17.0 $\mu$m emission features
and H$_2$ emission
in the northwest filaments; a strong 18.9 $\mu$m emission feature
both northeast and southeast of HD 200775;
and then a resurgence of the 16.4 $\mu$m feature and H$_2$
emission at the most southern positions.

Many of the observed positions on the sky coincide
to within $\sim$1 pixel ($\sim$5\arcsec)
for slits C and D. 
We overplot the independently
reduced spectra, at coincident spatial
positions, on Figures \ref{fig_b23} and \ref{fig_a6}.
These independently reduced spectra at similar (within
$\pm$5\arcsec) spatial positions are the best
assessment we can provide of the quality of our data,
because we are primarily limited by
systematics rather than by signal-to-noise.

\begin{figure*}
\figurenum{7}
\includegraphics[height=5.75in]
{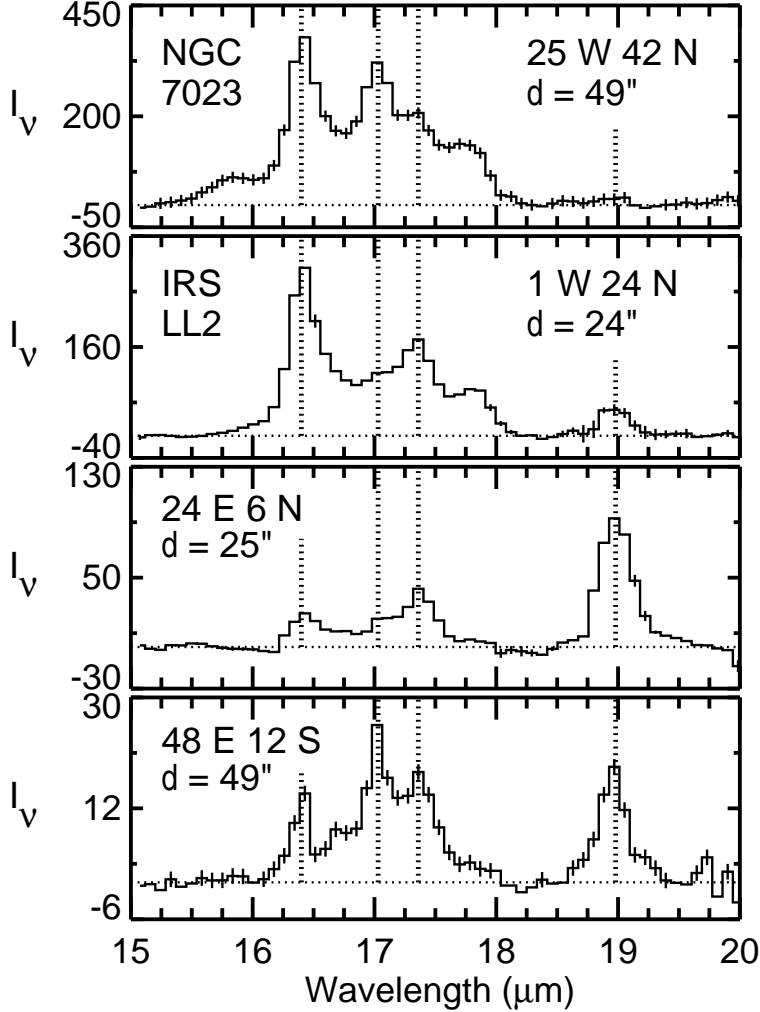}
\caption{
Intensity $I_\nu$ (MJy sr$^{-1}$),
after continuum subtraction,
vs. wavelength ($\mu$m), for
15 -- 20 $\mu$m LL2 
spectra of NGC 7023.
Note that the relative
strengths of the
16.4, 17.4, and 18.9 $\mu$m emission features
(wavelengths marked by {\it dotted 
lines})
depend more on nebular position (east or west)
than projected distance $d$ from the star.
North of the star
{\it (top two panels)},
at the nebular intensity peak,
spectra show strong 16.4 $\mu$m emission.
East of the star
{\it (bottom two panels)}, 
the 16.4 $\mu$m feature is much weaker,
and the 18.9 $\mu$m
feature appears strong by contrast.
The 17.4 $\mu$m feature is bright both 
north and east of the star.
The 0--0 S(1) H$_2$ line at
17.03 $\mu$m
(wavelength marked by {\it (dotted 
line)}
appears at $d$ = 49\arcsec.
For spectra at $d$ = 49\arcsec,
we plot $\pm$1-rms uncertainties calculated 
from the difference
between the spectrum and a
5-point moving boxcar average of the spectrum.
For spectra at $d$ = 24 -- 25\arcsec,
we plot $\pm$1-$\sigma$ uncertainties from
the average of two independent spectra
obtained at the same position.
}
\label{fig_newa}
\end{figure*}

We highlight some of our main results in Figure \ref{fig_newa}.
This figure illustrates two pairs of spectra 
at similar projected distances
($d$ = 24 -- 25\arcsec\ or $d$ = 49\arcsec) 
from HD 200775, 
but in different directions (east or west).
The continuum has been subtracted, by
fitting a parabola to points well-separated
from spectral emission features.

Figure \ref{fig_newa} demonstrates that
the LL2 spectra east and west of the star
have striking differences even at the same
projected distance from the star, and at
the same gas phase (neutral or molecular).
The two western spectra are in
the well-studied northwestern region of NGC 7023 containing
narrow, bright filaments marking the
PDR front there.
The two eastern spectra are typical of
the poorly-studied southeastern region of NGC 7023,
where we recently discovered 
the 18.9 $\mu$m interstellar emission feature 
\citepalias{WUS04}.

The 16.4 $\mu$m emission feature 
has the highest intensity (in MJy sr$^{-1}$) of any 
emission feature in the 
15 -- 20 $\mu$m spectra of NGC 7023 
(Figs. \ref{fig_b15}, \ref{fig_b17}, \ref{fig_b23}, 
\ref{fig_a6}, and \ref{fig_newa}).
The 16.4 $\mu$m feature emission, however,
with its strong spatial correspondence with the 
northwest filaments in NGC 7023, 
has much lower intensity east and southeast of the star.
Thus the 16.4 $\mu$m feature becomes
weaker than the 18.9 $\mu$m feature in the two
eastern spectra of Figure \ref{fig_newa}.

Figure \ref{fig_newa} shows that 
the 18.9 $\mu$m emission
feature is very prominent in the eastern spectra,
especially when
compared to its relative weakness in 
western spectra at similar $d$.
This difference is observed at the southeast
and northwest PDR
fronts, as traced by H$_2$ emission
($d$ = 49\arcsec),
and within the neutral H I gas ($d$ = 24 -- 25\arcsec).

Figure \ref{fig_newa} shows that
the 15.9 $\mu$m feature
is easily seen at $d$ = 49\arcsec\ NW,
hinted at in the spectrum of $d$ = 24\arcsec\ NW,
but is not detected at 
$d$ = 25\arcsec\ NE or $d$ = 49\arcsec\ SE.
We also conclude from Figure \ref{fig_newa} that
the 17.8 $\mu$m feature 
is observed only where the 16.4 $\mu$m emission feature
dominates the spectrum of NGC 7023: 
$d$ = 24 -- 49 \arcsec\ NW.
The 17.4 $\mu$m feature is apparent in all spectra
of Figure \ref{fig_newa}.
Its relative strength is difficult to estimate at this
spectral resolution, due to strong blending with H$_2$,
the 17.8 $\mu$m feature, and probably
at least one other feature (see below).

At the spectral resolution of LL2 ($R$ = 80 -- 128),
it is difficult to measure the relative
strengths of individual emission features,
particularly in the crowded 16 -- 18 $\mu$m spectral complex.
Still, there must be additional emission between the 16.4 $\mu$m
emission feature and the 17.4 $\mu$m feature in order to
explain the overall 16 -- 18 $\mu$m emission complex.
\citet{Smith06} fit this additional emission in LL2 spectra
with a broad (1.1 $\mu$m FWHM) feature at 17.0 $\mu$m.
We attempted a similar fit to this additional
emission in NGC 7023, at the resolution of LL2 spectra,
and were unable to find a unique fit.
We examine this broad emission, underlying the 15.9, 16.4,
17.4, and 17.8 $\mu$m features, further in the next section
using higher spectral resolution.

\subsection{Profile Fitting to Observed Spectra}
\label{gaussfit}

\begin{figure*}
\figurenum{8}
\includegraphics[width=6.50in]
{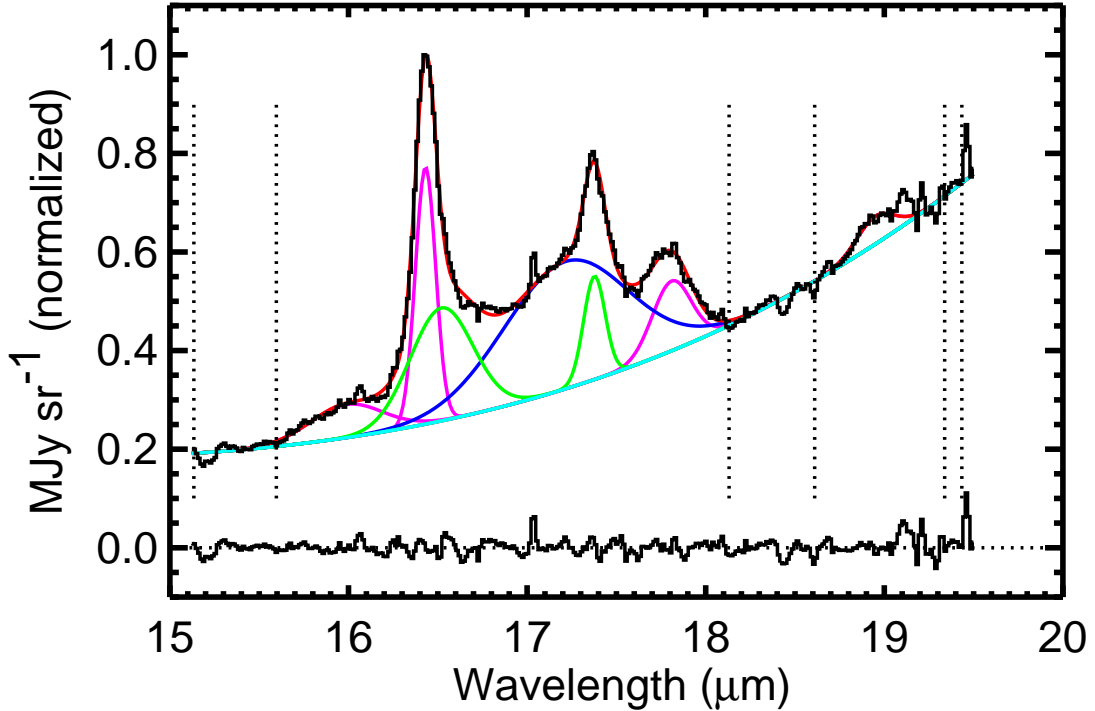}
\caption{
Normalized 
SH spectrum 
(15.1 -- 19.5 $\mu$m; $R$ = 600)
of Position B
(0\arcsec\ W 29\arcsec\ N; \citetalias{WUS04}) in NGC 7023 
({\it upper histogram}).
We measure the intensity $I_\nu$ (MJy sr$^{-1}$) 
from the entire entrance
slit (4\farcs7 $\times$ 11\farcs3).
We plot individual Gaussian profiles ({\it solid curves})
that have been fit to each feature (see text).
We mark spectral regions used to define the 
continuum ({\it vertical dotted lines}),
and illustrate the fitted parabolic
continuum ({\it solid line}).
We plot the final model spectrum
({\it solid curve}), which is a sum of the
fitted continuum and the individual Gaussian
profiles.
We plot the difference between our SH spectrum and the model
({\it lower histogram}) as an estimate of our noise.
}
\label{shgauss}
\end{figure*}

We model our SH spectrum, in order to quantify
the FWHM of each emission feature.
We use a program called LINER \citep{Pogge04}
to fit multiple blended Gaussians to the SH
spectrum we observe.
We illustrate our technique in Figure \ref{shgauss}.
We also ran fits with Lorentzian profiles \citep{Boulanger98},
which are similar to the Drude profiles used by
\citet{Smith06}; our conclusions do not depend on
whether a Gaussian or Lorentian profile is used.

First, we use a routine within LINER to interactively 
choose ranges of wavelengths for the continuum,
as shown in Figure \ref{shgauss}.
LINER requires a minimum of two wavelength ranges,
and a minimum of two wavelength points
per wavelength range.
The continuum for LINER is equal to a parabolic fit to the 
spectrum in these wavelength ranges.

Second, we use another routine within LINER
to make initial guesses at the height, width, and
center of each Gaussian that is used to model each feature
within the SH spectrum.
We do not fit profiles to features whose peak intensities
are less than 5\% of the continuum intensity,
taking this as the limit for possible calibration
artifacts \citep{Uchida05}.

Third, we use an interactive plotting tool within LINER to 
refine our guesses for the Gaussian parameters of
each feature.
This routine plots the
individual Gaussians and the 
sum of the multi-Gaussian fit,
resulting from our refined guesses, plotted over the data. 
This routine also plots the residuals
between the data and the fit 
to highlight where the model needs adjustment.
This routine allows us to modify our refined guesses,
by adding or subtracting features,
or by changing a feature's central wavelength, peak intensity, or FWHM,
in response to the quality of the model fit and its residuals.
It is during this step that broader features, such
as the 
17.0 $\mu$m feature, 
are better defined.
We iterate within this routine, modifying the
refined guesses, until we have fit the observed spectrum 
by eye as well as we can.

Fourth, we use a routine within LINER which 
uses standard numerical techniques
to search for best fit parameters for each feature (modeled as
a Gaussian) that minimizes
the difference between the entire model fit and the observed
spectrum,
based on the refined guesses we provide.
If the routine does not converge on a solution that is similar
to our refined guesses, we repeat the process, modifying our
initial and refined guesses, until we 
and the software concur on the fit.

We show the resulting fit, using this technique, 
for our SH spectrum 
in Figure \ref{shgauss}.
We illustrate the wavelength ranges for the continuum points,
the derived continuum, the
individual Gaussians, the sum of the multi-Gaussian fit,
and the residuals plotted together with the
original spectrum.
The residuals are another measure of the uncertainty
in our SH spectrum. 

We measure the central wavelength and
full width at half maximum (FWHM) 
of emission features in our SH spectrum ($R$ = 600).
We list these derived properties in Table 1.
All emission features are resolved except for H$_2$.

The wavelength calibration is uncertain for our SH spectrum,
because it was acquired during pre-nominal operations.
We therefore examine seven SH spectra $<$8\arcsec\ from
Position B, from
AOR key 3871232 (pid 28).
These spectra, obtained from a spectral map
with short integration times per position,
are noisier than our SH spectrum and so are not
reduced beyond the pipeline products or presented here.
These spectra, however, are sufficient to measure the
central wavelengths and FWHMs of all emission features
except the 18.9 $\mu$m emission feature.
For the 18.9 $\mu$m emission feature,
we select the ten spectra with the highest
signal-to-noise 18.9 $\mu$m emission feature, 
from the entire spectral map,
to measure the
central wavelength and FWHM of the 18.9 $\mu$m emission feature.
Nine of these spectra are 15 -- 42\arcsec\ southwest
of the star; one is 10\arcsec\ southeast of the star.

\clearpage
\begin{center}
\begin{tabular}{rrl}
\multicolumn{3}{c}{{\bf TABLE 1}}\\
\multicolumn{3}{c}{{\bf Observed Interstellar Emission 
Features at} $\lambda / \Delta \lambda$ {\bf = 600 in NGC 7023}}\\
\tableline
\tableline
\multicolumn{1}{c}{$\lambda_c^a$}&
\multicolumn{1}{c}{FWHM$^{\it a}$}\\
\multicolumn{1}{c}{($\mu$m)}&
\multicolumn{1}{c}{($\mu$m)}&
\multicolumn{1}{l}{References for Discovery$^{\it b}$}\\
\tableline
15.9 $\pm$ 0.1 &0.3 -- 0.6 & \citealt{Sturm00}\\
16.4 $\pm$ 0.1 &0.1 -- 0.2 & \citealt{Moutou00}; 
\citealt{VanKerckhoven00}; Sturm et \\
& & al. 2000\\
16.6 $\pm$ 0.1 &0.4 -- 0.9 & This paper \\
17.03 &0.021 & 0--0 S(1) H$_2$ \\
17.2 $\pm$ 0.1 &0.2 -- 0.8 & This paper \\
17.4 $\pm$ 0.1 &0.1 -- 0.2 & \citetalias{WUS04}\\
17.8 $\pm$ 0.1 &0.3 -- 0.4 & \citealt{Beintema96}\\
18.9 $\pm$ 0.1 &0.2 -- 0.3 & \citetalias{WUS04}\\
\tableline
\end{tabular}
\end{center}

\noindent
({\it a}) 
Central wavelength and FWHM
for all features measured from SH spectra
($\Delta \lambda$ = 0.021 $\mu$m).
The 0--0 S(1) H$_2$ line
at 17.03 $\mu$m, used to fix the
zero point of the wavelength calibration,
is unresolved.

\noindent
({\it b}) Reference for first detection of emission feature.

The very broad 16 -- 18 $\mu$m emission
plateau underneath most of the features in
Table 1 is not well fit by a single Gaussian
at $\sim$17.0 $\mu$m, as
found in the ISM previously at 
lower signal-to-noise
(\citealt{Beintema96}; \citealt{VanKerckhoven00})
or lower spectral resolution
(\citealt{Smith04}, \citealt{Smith06}).
Our best fit is with two broad features,
one at 16.6 $\pm$ 0.1 $\mu$m and another at 
at 17.2 $\pm$ 0.1 $\mu$m.
We find a best fit with two broad features both for Gaussian and
Lorentzian fits, although the central wavelengths
shift: 16.6 and 17.2 $\mu$m for a Gaussian
fit, or 16.8 and 17.1 $\mu$m for a Lorentzian fit.
The 16.6 $\mu$m feature might be a long 
wavelength wing to the 16.4 $\mu$m emission feature
rather than an independent emission feature,
analogous to the long wavelength wings of the 6.2 or
11.3 $\mu$m AEFs.

\section{Discussion}

\subsection{Laboratory comparison to NGC 7023 spectrum}

To prepare for the launch of {\it ISO},
\citet{Moutou96} published the 14 -- 40 $\mu$m 
laboratory spectra of neutral PAHs.
They considered a number of PAH mixes, and successfully predicted
the 16.4 $\mu$m feature in the ISM.  They did not, however,
predict the 18.9 $\mu$m emission feature.
They also did not predict
any of the other emission features observed in NGC 7023 
and elsewhere in the ISM,
such as the 15.9, 17.4, or 17.8 $\mu$m features, or the
broad 17.0 $\mu$m feature which we fit best by two 
features at 16.6 and 17.2 $\mu$m.

\citet{Peeters04} construct the spectra of two laboratory mixes of
PAHs, from the extensive PAH spectral database at NASA Ames
Research Center.
They compare these two mixes to {\it ISO}-SWS spectra of
two different AEF sources, and find good fits.
One source is dominated by the broad 17.0 $\mu$m feature,
with no other strong emission features.
The other source, \objectname{CD~$-42$~11721},
is dominated by the 16.4 $\mu$m feature,
as are many of our NGC 7023 spectra.

\begin{figure*}
\figurenum{9}
\includegraphics[width=6.50in]
{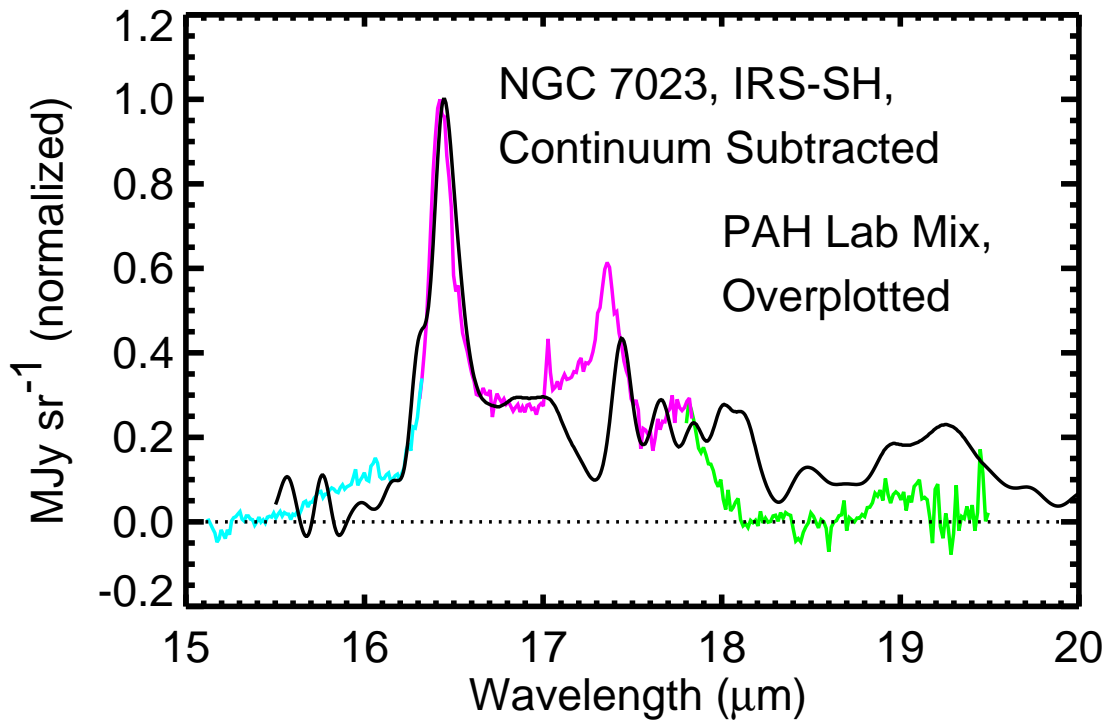}
\caption{
Our continuum-subtracted SH spectrum
of NGC 7023 (15.1 -- 19.5 $\mu$m; $R$ = 600; see
\citetalias{WUS04} and
Fig. \ref{fig_sh} for details).
We illustrate three overlapping SH orders, in contrasting colors.
Overplotted {\it (black curve)} is a laboratory spectrum from
\citet{Peeters04}, chosen to be a good fit to the {\it ISO}-SWS
spectrum of CD $-42$ 11721. 
NGC 7023 has an
{\it ISO}-SWS spectrum similar to that of 
CD $-42$ 11721 
\citep{Peeters04}.
}
\label{fig_lab}
\end{figure*}

\citet{Peeters04} overplot our SH spectrum 
(15.1 -- 19.5 $\mu$m; $R$ = 600; see 
\citetalias{WUS04} and
Fig. \ref{fig_sh})
of NGC 7023
on the {\it ISO}-SWS spectrum of CD~$-42$~11721,
to show the spectral similarities.
Both are dominated by the 16.4 $\mu$m feature.
They then plot a spectrum of a PAH laboratory mix on top of
their CD $-42$ 11721 spectrum, to show the
promising match in features.
But they do not
directly compare their spectrum of a PAH laboratory
mix to our SH spectrum of NGC 7023.
We show this comparison in Figure \ref{fig_lab}, with
the PAH laboratory spectrum courtesy of Peeters and
her collaborators.
The 16.4 $\mu$m feature in the laboratory mix is an excellent
match to the nebular 16.4 $\mu$m feature in NGC 7023.
The weaker substructure in the laboratory mix bears some
resemblance to the SH spectrum of NGC 7023.
Figure \ref{fig_lab} shows, however,
significant mismatches
when the laboratory and SH spectra are compared in detail.

\citet{Peeters04} suggest that the 18.9 $\mu$m feature
is due to ionized PAHs, based on 
fits to laboratory mixes of PAHs, but they
do not specifically search for a PAH
mix that would reproduce the spectrum of NGC 7023.
Furthermore, they do not explore a mix including
both neutral and ionized PAH species, as would be
expected for NGC 7023.
They plan a more detailed analysis of their laboratory
data in the future.
Their excellent match to the strong 16.4 $\mu$m emission feature,
and their modest success in producing weaker emission features
at 15 -- 20 $\mu$m, holds out promise that further
laboratory work may provide identifications
for some of the emission features at 15 -- 20 $\mu$m.

\subsection{Spatial variations in different emission features}

\begin{figure*}
\figurenum{10}
\includegraphics[height=6.20in]
{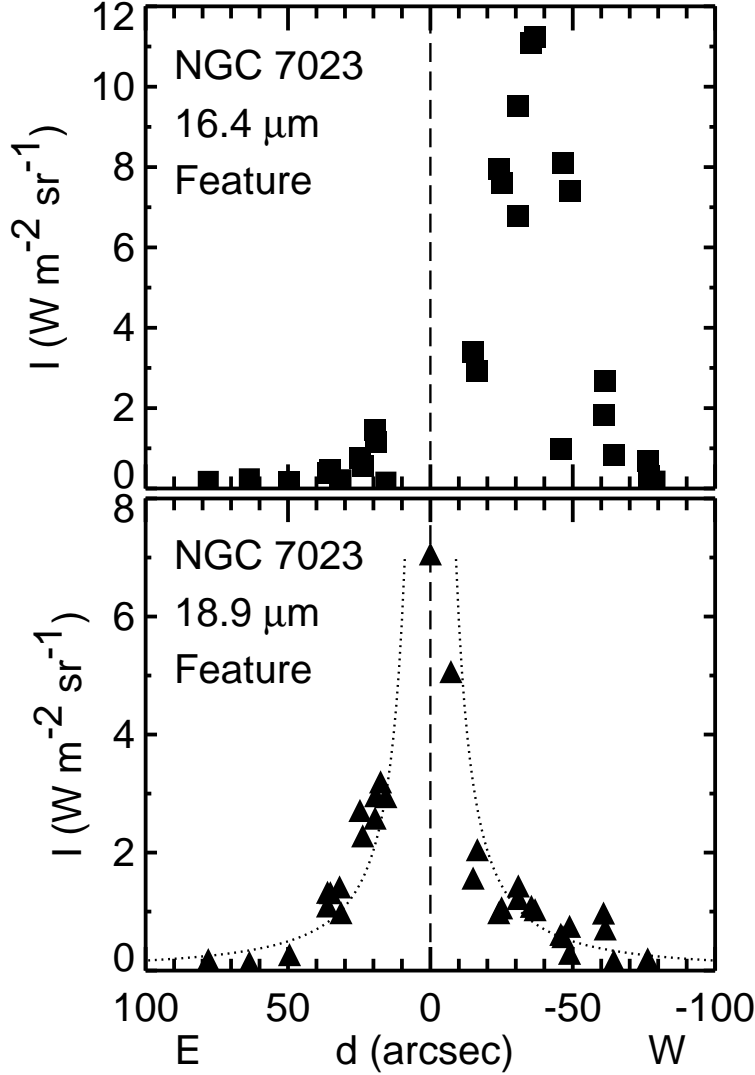}
\caption{The integrated feature intensities, $I$,
for the 16.4 $\mu$m
{\it (top; filled squares)} and 18.9 $\mu$m
{\it (bottom; filled triangles)} 
features, plotted vs. projected distance $d$ (arcsec) from HD 200775.
We plot values to the east (positive $d$) and west (negative $d$)
of the star separately.
The 16.4 $\mu$m feature intensity is strongly peaked in the west,
at a value of $d$ = $\sim$36\arcsec\ west.
The 18.9 $\mu$m feature intensity peaks on the star,
at $d$ = 0.
We fit a power-law, $I$ $\sim$ $d ^ {-1.54}$
{\it (dotted curve)}, to the 18.9 $\mu$m feature intensity.
We measure the
intensity 
(10$^{-20}$ W m$^{-2}$ sr$^{-1}$) extracted
in a 10\farcs2 $\times$ 15\farcs3 box
from LL2 ($R$ = 80 -- 128) spectra.
The angular resolution of each data point is 15\farcs3.
Statistical uncertainties are less than the point sizes;
the uncertainties are due to scatter among different
nebular locations at similar values of $d$.
}
\label{fig_164190}
\end{figure*}

We observe marked changes in spectral features
across NGC 7023 
(Figs. \ref{fig_b15}, \ref{fig_b17},
\ref{fig_b23}, \ref{fig_a6}, and \ref{fig_newa}),
which can be studied reliably using the long-slit
capabilities of the IRS on {\it Spitzer}.
In Figure \ref{fig_164190}, we have divided the 
16.4 and 18.9 $\mu$m feature intensity data
into nebular regions east and west of HD 200775, and
then plotted the spatial distribution of each.
The statistical uncertainties are less than the
sizes of the points in Figure \ref{fig_164190};
the uncertainty in this plot comes from the
intrinsic scatter between intensity values at
different nebular positions with similar projected
distances from the star.

Figure \ref{fig_164190} shows the 16.4 $\mu$m feature 
is strongly peaked 
$\sim$36\arcsec\ west of HD 200775.
This is near
the PDR front 
traced by H$_2$-emitting filaments
(\citealt{Lemaire96}; \citealt{An03}; \citealt{Witt06})
northwest of the star.

The 3.3, 6.2, and 11.3 $\mu$m AEF 
emission is strongly peaked on the
northwest filaments
(\citealt{Cesarsky96}; \citealt{An03}).
\citet{Moutou00} and \cite{VanKerckhoven00}
identify
the 16.4 $\mu$m feature with
C--C--C vibrational modes of neutral PAHs,
while the AEFs are widely attributed to PAHs.
The spatial distribution of the 16.4 $\mu$m emission feature
thus supports the current identification of it
with PAHs.

Figure \ref{fig_164190} also shows that, by contrast,
the 18.9 $\mu$m feature is symmetrically distributed
east and west of HD 200775, with a smooth fall-off
in intensity with $d$,
the projected distance from HD 200775.
We can fit a power-law to the distribution of the 18.9 $\mu$m
feature intensity, $I$, with $d$.  We find that $I$
$\sim$ $d ^ {-1.5}$.

We find the same results for both the 16.4 and 18.9 $\mu$m
features if the data
are instead split into northern and southern sections:
the 16.4 $\mu$m feature peaks in the north
at the PDR front, and the 18.9 $\mu$m
feature peaks symmetrically on the star.
The feature intensities in Figure \ref{fig_164190}
come from the area under a locally-defined
linear baseline, rather than the Gaussian fit
to the entire spectrum described in
\S \ref{gaussfit},
but the results
described in this section are unchanged if
feature intensities from Gaussian fits to the entire
spectra are analyzed instead.

A basic difference between the 16.4 $\mu$m feature 
and the 18.9 $\mu$m feature in NGC 7023 is that 
the 16.4 $\mu$m feature is bright in
the northwestern PDR, where the column density is
known to be high,
while the 18.9 $\mu$m
feature remains bright $\sim$30\arcsec\ east and south
of the star,
where it appears the column density
is low. 

The regions east and south of
the star appear, at least in projection, to lie inside
the fossil bipolar outflow observed in the
24 $\mu$m {\it Spitzer} MIPS emission
\citepalias{WUS04}.
The northwest PDR, by contrast, appears to be part of the
remnant molecular material surrounding the original
bipolar outflow.
It is possible that shock processing plays a role
in the creation of the 18.9 $\mu$m feature carrier.

\subsection{Identifying the 18.9 Micron Interstellar Emission Feature}

\subsubsection{Minerals}

The strength of the 18.9 $\mu$m feature near the star,
and its rapid falloff with $d$, 
leads us to consider whether the excitation
of this feature could be {\it equilibrium thermal emission}
from large grains,
rather than non-equilibrium emission
from tiny grains, PAHs, or other large molecules.
We can (with difficulty) construct
a model of equilibrium thermal emission that explains
the 18.9 $\mu$m feature itself, but cannot
construct a equilibrium thermal emission
model for the underlying 18.9 $\mu$m continuum.
In this case, the
observed 18.9 $\mu$m continuum would 
be due to non-equilibrium
emission, from tiny grains or large molecules.

In the equilibrium thermal emission model for the
18.9 $\mu$m feature, its carrier is large grains
of unknown composition.
We searched the literature for astrophysically important
minerals, and also searched the
Jena - St. Petersburg Database of Optical Constants at
http://www.astro.uni-jena.de/Laboratory/Database/jpdoc/index.html
\citep{Henning99}.
We can rule out the following substances as carriers of
the 18.9 $\mu$m feature:
graphite \citep{DL84},
amorphous carbon \citep{Colangeli95},
hydrogenated amorphous carbon \citep{OD88},
hydrocarbon nanoparticles \citep{Herlin98},
amorphous silicates (both olivines, Mg$_{2y}$Fe$_{2y-2}$SiO$_4$, 
and pyroxenes, Mg$_x$Fe$_{1-x}$SiO$_3$;
\citealt{Jaeger94, Dorschner95, Koike00}),
crystalline silicates (both olivines and pyroxenes;
\citealt{Koike93, Jaeger94, Koike00, Koike03}),
quartz (SiO$_2$; \citealt{KS94, Posch99}),
corundum ($\alpha-$Al$_2$O$_3$; \citealt{Posch99}), 
spinel (MgAl$_2$O$_4$; \citealt{Posch99, Fabian01}),
FeO \citep{Henning95, HM97}, 
MgO \citep{Henning95}, 
rutile (TiO$_2$; \citealt{Posch99}), 
SiC \citep{Mutschke99, HM01}, 
TiC \citep{HM01},
carbonates such as 
CaMg(CO$_3$)$_2$, CaCO$_3$, MgCO$_3$, FeCO$_3$ and CaFe(CO$_3$)$_3$
\citep{Kemper02},
FeS \citep{HM97}, 
MgS \citep{Hony02}, 
crystalline or amorphous SiS$_2$ \citep{Begemann96, Kraus97, HM97},
or silicon nitride (Si$_3$N$_4$; \citealt{Clement05}).

\subsubsection{Aromatic Molecules}

If the 18.9 $\mu$m feature is not due to 
emission from large grains in
thermal equilibrium, then it,
like the 16.4 $\mu$m feature,
is due to {\it non-equilibrium emission}
from nanoparticles: tiny grains, PAHs, or other large molecules.
Here we discuss one main class of large molecules:
aromatics such as PAHs.

\citet{Rapacioli05} combine {\it ISO}-CAM 
circular variable filter (CVF)
images of NGC 7023
with laboratory and theoretical spectra of neutral and
ionized PAHs, and analyze these 
with a decomposition technique,
to construct what they call images of the ionized PAHs (PAH$^+$) and
neutral PAHs (PAH$^0$) in NGC 7023.
They measure the strength of PAH$^+$ and PAH$^0$ emission
along a line from the star, through the northwest
filaments, into the molecular cloud.
Their results show that PAH$^+$ does not peak on
the star, but rather peaks between the star and
the northwest filaments, at $d$ = $\sim$24\arcsec\ from the star.
They find that PAH$^0$ peaks at $d$ = $\sim$42\arcsec\
from the star, in the northwest filaments.

\citet{Flagey06}
use {\it ISO}-CAM CVF images plus {\it Spitzer} IRAC images
to derive improved images
for PAH$^+$ and PAH$^0$ in NGC 7023,
with a similar decomposition technique.
Their images show the same results as those of \citet{Rapacioli05}.
\citet{Flagey06} find that PAH$^+$ peaks 
midway between the star and the northwest filaments,
at $d$ = $\sim$22\arcsec, not on the star itself.
\citet{Flagey06} observe PAH$^0$ to peak on the northwest filaments,
at $d$ = $\sim$40\arcsec.

\begin{figure*}
\figurenum{11}
\includegraphics[height=6.65in]
{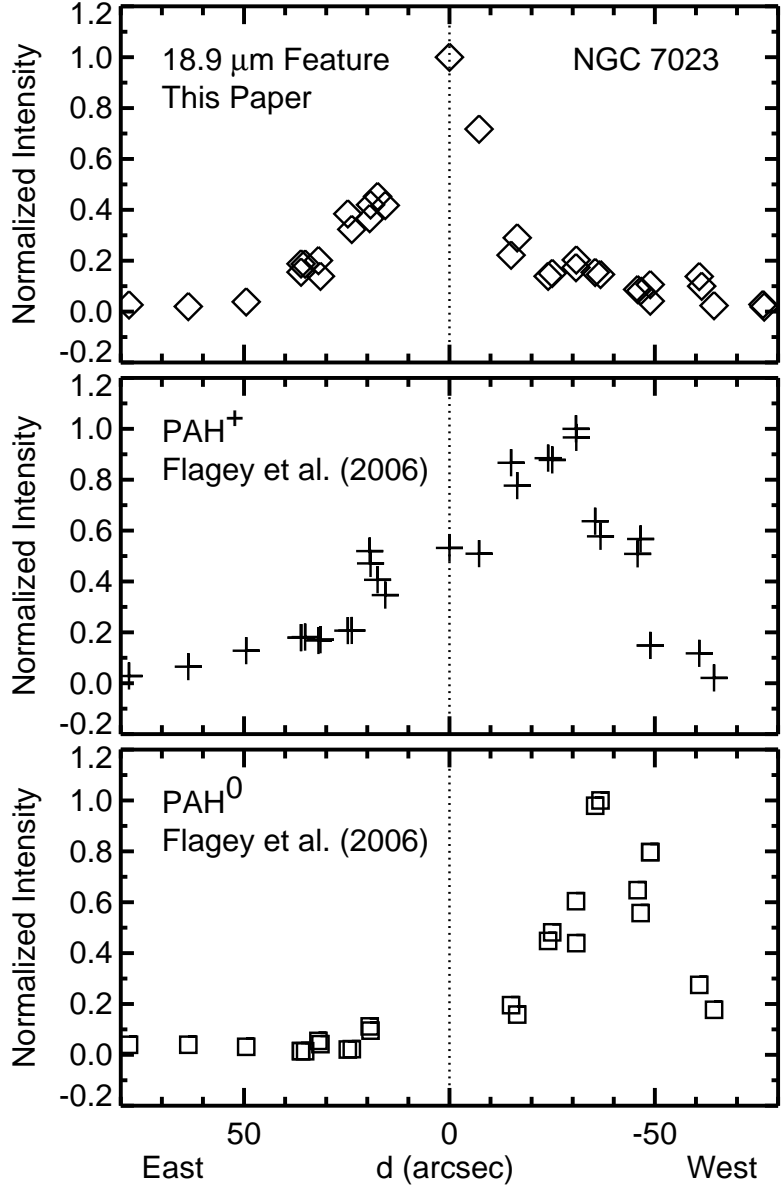}
\caption{
Normalized intensity versus projected distance $d$ (arcsec) from HD 200775,
for the 18.9 $\mu$m emission feature
({\it diamonds}; this paper), 
PAH$^+$ ({\it plus signs}; \citealt{Flagey06}),
and 
PAH$^0$ ({\it squares}; \citealt{Flagey06}).
We plot values to the east (positive $d$) and west (negative $d$)
of the star separately.
The 18.9 $\mu$m emission feature peaks on
HD 200775, at $d$ = 0;
PAH$^+$ peaks between the star and the
northwest filaments, at $d$ = $\sim$22\arcsec\ west;
PAH$^0$ peaks on the northwest filaments,
at $d$ = $\sim$40\arcsec\ west.
The angular resolution of each data point is 15\farcs3,
6\arcsec, and 6\arcsec\ for 18.9 $\mu$m, PAH$^+$,
and PAH$^0$, respectively.
}
\label{fig_flagey}
\end{figure*}

We compare the spatial distribution of our 18.9 $\mu$m emission
feature intensity to the \citet{Flagey06} results in
Figure 
\ref{fig_flagey}. 
We calculate PAH$^+$ and PAH$^0$ relative intensities,
from images kindly provided to us by
\citet{Flagey06},
at the same spatial positions at which we
measure the 18.9 $\mu$m feature intensity.
Figure
\ref{fig_flagey} vividly illustrates how unique the 18.9 $\mu$m
feature distribution is compared to the distribution
of neutral and singly ionized PAHs inferred by
\citet{Flagey06}.
Figure 
\ref{fig_flagey} 
shows that
the peak 18.9 $\mu$m feature emission is centered
at $d$ = 0, 
in contrast to the regions of peak PAH$^+$ 
($d$ = $\sim$22\arcsec\ west) 
and PAH$^0$
($d$ = $\sim$40\arcsec\ west) emission.
This is true even if one excludes the 18.9 $\mu$m data point
at the position of HD 200775.

Figure 
\ref{fig_flagey} 
clearly demonstrates that
the distribution of intensity for
the 18.9 $\mu$m feature we observe is spatially quite
distinct from the distributions for PAH$^+$ and PAH$^0$
derived by \citet{Flagey06}.
A similar plot of the
\citet{Rapacioli05} PAH$^+$ and PAH$^0$ 
data yields the same result.
PAH$^+$ and PAH$^0$ are appealing carriers for the
18.9 $\mu$m feature, because of their likely ubiquity
in space and the probability that laboratory data and
theoretical modeling of mixes of their spectra at
15 -- 20 $\mu$m will be published in the near future.
If the analyses of \citet{Rapacioli05} and \citet{Flagey06}
are correct, however, concerning the location of PAH$^+$ and PAH$^0$
in NGC 7023, then PAH$^+$ and PAH$^0$
would both be unlikely carriers for the 18.9 $\mu$m feature.

\citet{Witt06}
observe the 
the Extended Red Emission (ERE) in NGC 7023 
to be bright in the northwestern filaments.
They show that the ERE is coincident in most cases
with the H$_2$ emission in the filaments.
The 16.4 $\mu$m feature peaks in these
filaments (Fig. \ref{fig_164190}),
as do the AEFs (\citealt{Cesarsky96}; \citealt{An03}).
This argues against the ERE carrier and the 18.9 $\mu$m
carrier being the same.
\citet{Witt06}
propose that PAH dications (PAH$^{++}$) are the carriers of the ERE.
This is a tentative argument against
PAH$^{++}$ being the carriers for the 18.9 $\mu$m emission
feature, but the PAH$^{++}$ identification 
model for the ERE \citep{Witt06} 
needs to be confirmed.

If a mix of neutral and ionized PAHs does not provide
a match to our NGC 7023 spectra, then
there are many other PAH variants to explore.
A long list of references is given in \citet{Halasinksi05}, including
(among others) PAHs with side groups,
nitrogen-substituted PAHs, dehydrogenated PAHs and protonated PAHs.

\subsubsection{Fullerenes}

The 18.9 $\mu$m feature has a spatial distribution
quite distinct from that of the 16.4 $\mu$m feature in NGC 7023.
Figure \ref{fig_newa} illustrates that
spectra west of the star are dominated by the 16.4 $\mu$m
feature (likely due to PAHs), while 
spectra east of the star are dominated
by the 18.9 $\mu$m feature.
Figure \ref{fig_164190} shows that
the 18.9 $\mu$m feature peaks on the star, with an
intensity proportional to $d ^ {-1.5}$;
the 16.4 $\mu$m feature shows no dependence on $d$.
This spatial distinction between the 18.9 and
16.4 $\mu$m features suggests that
two different species may be needed, rather than
two ionization states of similar species.

\citet{Jura04} has suggested that the 17.4 and 18.9 $\mu$m
features in NGC 7023 are due to neutral C$_{60}$.
C$_{60}$ would emit by non-equilibrium emission
in NGC 7023 \citep{Moutou99}.
\citet{Kwok99} detect narrow features at 17.85 and 18.90 $\mu$m
in the C-rich protoplanetary nebula
\objectname[]{IRAS~07135+1005}.
They tentatively detect a third narrow feature at
19.15 $\mu$m.
\citet{Kwok99} note that 
neutral C$_{60}$ has 
four fundamental vibrational frequencies,
at 7.1, 8.6, 17.5 and 19.0 $\mu$m \citep{Frum91}.
They stop short of identifying the features observed in
IRAS 07135+1005 as due to C$_{60}$, however.
The emission feature in IRAS 07135+1005 at 17.85 $\mu$m is not
a good wavelength match for the C$_{60}$
17.5 $\mu$m line.
Furthermore, \citet{Kwok99} do not detect the 8.6 $\mu$m C$_{60}$
feature in IRAS 07135+1005 (they are unable to search
for the 7.1 $\mu$m line).

We observe emission features at
17.4 and 18.9 $\mu$m, together, to the east and southeast of the star.
These two features are a better wavelength
match to the 17.5 and 19.0 $\mu$m lines of
neutral C$_{60}$.
This suggests that
these two features might be due to C$_{60}$.
The 17.4 $\mu$m feature, however, is also observed 
northwest of the star,
where the 16.4 feature peaks and the
18.9 $\mu$m feature is weak.  
This casts doubt on attributing the 17.4 $\mu$m feature,
at least, to C$_{60}$.

Two possibilities exist to explain why the 17.4 $\mu$m
feature is observed in both east and west of the star,
while the 18.9 $\mu$m feature is not.
First, the 17.4 $\mu$m feature could be unrelated to the
18.9 $\mu$m feature.
In this case, neither feature is due to C$_{60}$,
and the 17.4 $\mu$m feature has a spatial distribution
which is distinct from both the 16.4 and the 18.9 $\mu$m
features.
Second, the 17.4 $\mu$m feature could be a blend (at the
lower resolution of our LL2 spectra) of two features.
In this case, one contributor to the 17.4 $\mu$m
intensity, perhaps C$_{60}$, follows
the 18.9 $\mu$m feature, while the other contributor to the 
17.4 $\mu$m intensity,
probably PAHs, follows the 16.4 $\mu$m feature.
These two possibilities might be distinguished by
comparing higher resolution 15 -- 20 $\mu$m spectra
east and west of the star.

Another way to test the possibility that the 17.4 and 18.9 $\mu$m
features are from C$_{60}$ is to search for
emission from its other two vibrational modes
at 7.1 and 8.6 $\mu$m \citep{Frum91}.
\citet{Moutou99} searched for the 7.1 and 8.6 $\mu$m lines of C$_{60}$
at the nebular peak of NGC 7023, near the northwest PDR,
and placed strong upper limits on C$_{60}$. 
The 18.9 $\mu$m feature is weak in this northwest region, 
however, so this
does not rule out C$_{60}$ as an identification
for the 18.9 $\mu$m feature.
By obtaining 7 -- 9 $\mu$m 
spectra east and southeast of the star,
where the 18.9 $\mu$m feature is observed,
one could search for the 7.1 and 8.6 $\mu$m lines of C$_{60}$,
to directly test the hypothesis that the 18.9 $\mu$m
feature (and part of the 17.4 $\mu$m feature intensity) is due
to C$_{60}$.

\citet{Moutou99} use the models of
\citet{BT94} to show that
fullerenes in NGC 7023 should be divided between neutral
C$_{60}$ (35\%) and ionized C$_{60}^+$ (50\%).
\citet{Moutou99} also put
stringent upper limits on C$_{60}^+$ at the
NGC 7023 nebular peak, from the absence
of the 7.1 and 7.5 $\mu$m vibrational modes of C$_{60}^+$
\citep{Fulara93}.
\citet{Moutou99}, however, did not 
search for these emission lines east or southeast of the star.
Spectra of NGC 7023 at 7 -- 9 $\mu$m, 
at these positions where the 18.9 $\mu$m
feature is bright, would simultaneously search for the
7.1 and 7.5 $\mu$m lines of C$_{60}^+$ as well as the
7.1 and 8.6 $\mu$m lines of C$_{60}$.
No laboratory spectra at 15 -- 20 $\mu$m 
exist for C$_{60}^+$, so it is not known
where its longer wavelength lines might be.
If C$_{60}$ is present in NGC 7023, however, then C$_{60}^+$ should
also be present at a similar abundance.
Laboratory spectra at 10 -- 40 $\mu$m of C$_{60}^+$ are needed,
to measure the wavelengths of all four infrared lines,
and to measure their relative strengths.

C$_{70}$ has lines at 17.4, 18.1, and 19.0 $\mu$m \citep{Nemes94}.
The lines at 18.1 and 19.0 $\mu$m have equal intensity in
gas-phase C$_{70}$ emission \citep{Nemes94}.
The absence of a detectable 18.1 $\mu$m feature in NGC 7023,
particularly in the regions where the 18.9 $\mu$m feature
is observed to be strong, makes C$_{70}$ an unlikely identification for
the 17.4 or 18.9 $\mu$m features.

\subsubsection{Diamonds}

Many authors suggest that
interstellar nanodiamond grains, similar to 
pre-solar nanodiamonds extracted
from meteorites,
can explain some
infrared emission features seen in the ISM
or circumstellar environments 
(\citealt{Koike95};
\citealt{Hill97};
\citealt{Hill98};
\citealt{Guillois99};
\citealt{VanKerckhoven02}).
Pre-solar meteoritic nanodiamonds
have a median radius of 1.3 nm,
and thus can radiate by non-equilibrium thermal emission
in the ISM
\citep{Koike95}.

Laboratory spectra of pre-solar nanodiamonds 
have problems with surface contamination
of the nanodiamonds, which adds strong
features that can mask the weak features of
diamond.
Pure diamond itself has no infrared-active lines
at 15 -- 20 $\mu$m, but impurities and
crystal defects break the symmetry and
allow lines to appear in the diamond spectrum.
Laboratory spectra of pre-solar nanodiamonds (from the
Allende, Murchison, and Orgueil meteorites),
natural (terrestrial) diamonds, and
synthetic diamonds
(\citealt{Lewis89}; 
\citealt{Colangeli94};
\citealt{Koike95};
\citealt{Mutschke95};
\citealt{Hill97}; 
\citealt{Andersen98};
\citealt{Braatz00}; 
\citealt{Mutschke04}) 
show no features at 18.9 $\mu$m and thus
rule out nanodiamonds
as identifications for the 18.9 $\mu$m feature.

\citet{Oomens06} argue that
the identification of interstellar diamonds in meteorites
make it worth considering smaller
diamonds, such as diamondoids, as possible ISM components.
Diamondoids, whose surfaces are fully
hydrogenated, are the smallest diamond molecules.
These diamond molecules can have a varying number of diamond cages,
such as one (adamantane; C$_{10}$H$_{16}$),
two (diamantane; C$_{14}$H$_{20}$), or three 
(triamantane; C$_{18}$H$_{24}$).
The 3 -- 15 $\mu$m spectra of diamondoids with one through
six diamond cages have been measured in the laboratory by
\citet{Oomens06}.
No laboratory data exist at 15 -- 20 $\mu$m.
\citet{Lu05} predict vibrational frequencies at 15 -- 20 $\mu$m for
diamondoids with one through ten diamond cages.
They do not predict any features at 18.9 $\mu$m
for the first five diamond cages.
\citet{Pirali06}, who predicts vibrational frequencies 
at 15 -- 20 $\mu$m for
the same diamondoids measured by \citet{Oomens06},
predicts a 19.1 $\mu$m line for
[121]tetramantane. 
This 19.1 $\mu$m line is accompanied by a
much stronger 16.3 $\mu$m line, however, which 
does not explain
the strong 18.9 $\mu$m and
faint 16.4 $\mu$m feature emission to the east and southeast of the star.
Laboratory spectra of diamondoids at 15 -- 20 $\mu$m
are needed,
but it looks unlikely that diamondoids can 
explain the 18.9 $\mu$m feature.

\subsection{Further work}

Further observations, and further
laboratory work, are needed to better identify the 18.9 $\mu$m
emission feature in NGC 7023, as well as other
emission features at 15 -- 20 $\mu$m.

Further observations are needed
where the 18.9 $\mu$m feature is bright in NGC 7023,
east and southeast of the star, particularly
high-resolution spectra at 7 -- 9 $\mu$m and 15 -- 20 $\mu$m.
Spectra 
should be obtained with enough spectral resolution
to cleanly measure the strength of the 17.4 $\mu$m
feature (LL2 spectra are too low resolution),
and compared east and west of the star,
to search for any blending of the 17.4 $\mu$m feature.
This tests the hypothesis that the 18.9 $\mu$m
feature (and part of the 17.4 $\mu$m feature intensity) is due to
C$_{60}$.
New 7 -- 9 $\mu$m spectra in regions where the 
18.9 $\mu$m feature dominates the spectrum are
needed to test the C$_{60}$ identification, 
by searching for the 7.1 and 8.6 $\mu$m lines of C$_{60}$.
This will also simultaneously search for the 
7.1 and 7.5 $\mu$m lines of C$_{60}^+$.
High spectral resolution ($R$ $\sim$ 2000)
is important at 7 -- 9 $\mu$m because the
C$_{60}$ features are superposed
on the very strong 7.7 $\mu$m AEF and
are expected to be weak.

The firm detection of the 18.9 $\mu$m feature in
only two out of a sample of 27 H II nuclei galaxies 
\citep{Smith06}
demonstrates that this feature exists outside
NGC 7023, but it is not common.
Detailed observations of the phenomenology for
all of the 15 -- 20 $\mu$m features,
but especially the 18.9 $\mu$m
feature, are important to obtain
in sources other than NGC 7023.

Published 15 -- 20 $\mu$m laboratory spectra of PAHs, 
fullerenes (especially
C$_{60}^+$), and nanodiamonds/diamondoids
are essential; these species, so far,
are the leading candidates for the 18.9 $\mu$m feature.
But the spatial dependence of the 18.9 $\mu$m
feature intensity $I$ in NGC 7023, 
with $I$ $\sim$ $d^{-1.5}$ (Fig. \ref{fig_164190}),
suggests that ionized and photochemically
altered forms of these materials
should be especially pursued.

Alternatively, the 18.9 $\mu$m
feature carrier could be a photodissociation or shock product,
or a species only able to exist or survive near the star.
Theory aimed at understanding the spatial distribution
of the 18.9 $\mu$m feature intensity in NGC 7023
may be able to predict new ISM components for which
15 -- 20 $\mu$m spectra laboratory spectra could be
obtained, and compared with 15 -- 20 $\mu$m
spectra of NGC 7023.

We are interested in identifying all the features
at 15 -- 20 $\mu$m, not merely the 18.9 $\mu$m feature.
In this context,
laboratory 15 -- 20 $\mu$m spectra should be obtained of
PAH variants which have proven fruitful for studies of the AEFs,
such as PAHs with side groups,
nitrogen-substituted PAHs, protonated PAHs,
or dehydrogenated PAHs
\citep{Halasinksi05}. 

A final possibility is that nanoparticles (radius $\sim$1 nm) of minerals
thought or known to be present in the ISM, such as
silicates or SiC, might have spectra that differ from
those of the solid-state mineral.
Such nanoparticles would emit by non-equilibrium
emission in NGC 7023.
Infrared gas-phase emission spectra of 1 -- 2 nm radius nanoparticles,
when technically feasible in the
laboratory, would be an important
area of exploration for ISM studies.

\section{Conclusions}

We observe the 15 -- 20 $\mu$m spectra of
the photodissociation region (PDR) and reflection nebula
NGC 7023.
We analyze four LL2 slit pointing positions in NGC 7023,
including a re-reduction of the one LL2 slit pointing 
position presented in \citetalias{WUS04}.
We extract spectra at eight spatial positions along each LL2 long-slit.
We fit blends of Gaussian profiles to our observed
SH spectrum from \citetalias{WUS04}, to derive 
central wavelengths and FWHMs
for each interstellar emission feature.

We observe spectral emission features at 15.9, 16.4,
17.4, 17.8, and 18.9 $\mu$m.
All of these features are resolved in our SH spectrum.
We observe the broad 17.0 $\mu$m emission feature
first discovered by \citet{Beintema96}, and
later studied by 
\citet{VanKerckhoven00}, \citet{Smith04}, and \citet{Smith06}.
We find, however, that it is better fit at higher spectral
resolution by two features at 16.6 and 17.2 $\mu$m.
We observe 0--0 S(1) H$_2$ emission, particularly
at the filaments northwest of the star which mark the
edge of the PDR there, and again at the
PDR southwest of the star.

The 16.4 $\mu$m feature is the brightest emission
feature northwest of the star, and
peaks in the northwest filaments.
Southeast of the star, by contrast,
the 18.9 $\mu$m feature is the brightest
emission feature. 
The 18.9 $\mu$m feature
peaks on the
star, with its intensity falling off with projected
distance $d$ from the star as $d^{-1.5}$.
The 17.4 $\mu$m feature is present 
both northwest and southeast of the star.

\citet{Moutou00} and \citet{VanKerckhoven00}
identify
the 16.4 $\mu$m feature with
C--C--C vibrational transitions of neutral PAHs.
We observe that the 16.4 $\mu$m emission feature
peaks $\sim$36\arcsec\ northwest of the star; 
this is in the PDR 
traced by H$_2$-emitting filaments 
(\citealt{Lemaire96}; \citealt{An03}; \citealt{Witt06}).
The 3.3, 6.2, and 11.3 $\mu$m AEFs
also peak in the northwest filaments
(\citealt{Cesarsky96}; \citealt{An03}).
The AEFs are widely attributed to PAHs,
thus supporting the current laboratory identification of PAHs
with the 16.4 $\mu$m feature.

\citet{Peeters04}
propose the 18.9 $\mu$m
feature is due to ionized PAHs.
\citet{Rapacioli05} and \citet{Flagey06}
derive images of 
singly ionized PAHs (PAH$^+$) 
and neutral PAHs (PAH$^0$)
in NGC 7023.
PAH$^+$ peaks at
$d$ = 20 -- 22\arcsec, between the star and the 
northwest filaments; 
PAH$^0$ peaks at $d$ = 40 -- 42\arcsec, 
on the northwest filaments 
(\citealt{Rapacioli05}; \citealt{Flagey06}).
We observe the 18.9 $\mu$m feature to peak at $d$ = 0\arcsec,
and to be symmetrically distributed around the star with no
peak at the northwest filaments.
If the analyses of
\citet{Rapacioli05} and \citet{Flagey06}
are correct, then 
PAH$^+$ and PAH$^0$ are unlikely as
identifications for the 18.9 $\mu$m feature.
\citet{Witt06} propose that PAH dications (PAH$^{++}$) are the 
carriers of the Extended Red Emission (ERE) in NGC 7023,
but they observe the
ERE to be bright in the northwest filaments.
If their identification of the ERE with PAH$^{++}$ is
correct, then that would make PAH$^{++}$ unlikely as an identification
for the 18.9 $\mu$m feature.
Other PAH variants remain to be explored.

\citet{Jura04} notes that
neutral C$_{60}$ has features at 17.5 and 19.0 $\mu$m,
while we observe emission features in NGC 7023 
at 17.4 and 18.9 $\mu$m.
Southeast of the star, we see these features 
together, with the 18.9 $\mu$m feature always
stronger than the 17.4 $\mu$m feature.
Northwest of the star, however,
we observe the 17.4 $\mu$m feature alone or much
stronger than the 18.9 $\mu$m feature.
Either the 17.4 and 18.9 $\mu$m features are not due
to C$_{60}$, or the 17.4 $\mu$m feature is due to
C$_{60}$ plus a second 17.4 $\mu$m carrier with a different
spatial distribution.

\citet{Moutou99} suggest that
$\sim$50\% of the C$_{60}$ in NGC 7023 should
be singly ionized in NGC 7023.
The wavelengths for the two longer wavelength infrared active 
C$_{60}^+$ modes
are not known; they are likely to fall near
the wavelengths for C$_{60}$ (i.e. 17.5 and 19.0 $\mu$m),
but need to be measured in the laboratory.
The 7 --  9 $\mu$m lines of 
C$_{60}$ and C$_{60}^+$ have been
searched for in NGC 7023, but not yet southeast
of the star where the 18.9 $\mu$m feature
is bright; these observations also need to be made.

We search the literature for other identifications for
the 18.9 $\mu$m feature.
We do not find a match among all other minerals that
have been, to our knowledge, been proposed as components
of the ISM and for which we were able to locate 15 -- 20 $\mu$m
laboratory spectra.
In particular, we rule out amorphous silicates, crystalline
silicates, amorphous carbon, graphite, SiC, corundum, and
diamonds.

We desire a better understanding of 
the 15.9, broad 17.0, 17.4, 17.8, and 18.9 $\mu$m
emission features we observe in NGC 7023. 
This awaits new and better
laboratory data on 
PAH variants,
C$_{60}^+$, 
other fullerenes, nanodiamonds, diamondoids, and related species.
Laboratory spectra, and ISM theory as a
guide, are essential to sort through the possible identifications
of these features.
Detailed observations of these features in sources other
than NGC 7023 will determine how widespread each feature
is, and clarify whether the same phenomenology is
observed elsewhere.

\acknowledgments
We thank the teams and people whose dedicated
work made the {\it Spitzer} Space Telescope a reality
(see \citealt{SIRTF}).
We greatly appreciate the generosity of 
S. Casey,
N. Flagey,
J. Oomens,
E. Peeters,
and O. Pirali,
who shared electronic versions of
their results with us.
We acknowledge helpful interchanges with
L. J. Allamandola,
D. J. Ennis,
F. Kemper,
O. Pirali,
R. W. Pogge,
J. D. Smith,
and anonymous reviewers.
This work is based on observations made with 
the {\it Spitzer} Space Telescope, which is operated by the Jet 
Propulsion Laboratory, California Institute of Technology 
under NASA contract 1407. Support for this work was provided 
by NASA through Contract Number 1257184 issued by JPL/Caltech,
and by NSF grant AST 02-06331.
KS thanks the NASA Faculty Fellowship Program for financial
support, and R. Pogge for assistance with using LINER.
This research has made use of NASA's Astrophysics Data System;
the SIMBAD database,
operated at CDS, Strasbourg, France;
and data products from the Two Micron 
All Sky Survey, which is a joint project of the University of 
Massachusetts and the Infrared Processing and Analysis 
Center/California Institute of Technology, funded by the 
National Aeronautics and Space Administration and the National 
Science Foundation.

{\it Facility:} \facility{Spitzer (IRS)}

\clearpage

\clearpage
  
\clearpage

\clearpage

\clearpage

\clearpage

\clearpage

\clearpage

\clearpage

\clearpage

\clearpage

\end{document}